\documentclass[aps,prl,twocolumn,superscriptaddress]{revtex4-1}
\usepackage{graphicx,amsmath,amssymb,bm}
\usepackage[
   bookmarks=true,
   colorlinks=true,
   hyperindex=true,
   citecolor=blue,
   linkcolor=blue,
   urlcolor=red
]{hyperref}

\begin{document}

\title{Moir\'e-Enabled Recursive Phase Unwrapping for Differential Atom Interferometry}

\author{Junjie Jiang}
\altaffiliation{These authors contributed equally to this work.}
\affiliation{Innovation Academy for Precision Measurement Science and Technology, Chinese Academy of Sciences, Wuhan 430071, China}
\affiliation{School of Physical Sciences, University of Chinese Academy of Sciences, Beijing 100049, China}

\author{Sitong Yan}
\altaffiliation{These authors contributed equally to this work.}
\affiliation{Innovation Academy for Precision Measurement Science and Technology, Chinese Academy of Sciences, Wuhan 430071, China}

\author{Jiaming Wei}
\affiliation{Innovation Academy for Precision Measurement Science and Technology, Chinese Academy of Sciences, Wuhan 430071, China}
\affiliation{School of Physical Sciences, University of Chinese Academy of Sciences, Beijing 100049, China}

\author{Chuan He}
\affiliation{Innovation Academy for Precision Measurement Science and Technology, Chinese Academy of Sciences, Wuhan 430071, China}

\author{Xi Chen}
\affiliation{Innovation Academy for Precision Measurement Science and Technology, Chinese Academy of Sciences, Wuhan 430071, China}
\affiliation{Hefei National Laboratory, Hefei 230088, China}
\affiliation{Wuhan Institute of Quantum Technology, Wuhan 430206, China}

\author{Lin Zhou}
\email{lzhou@wipm.ac.cn}
\affiliation{Innovation Academy for Precision Measurement Science and Technology, Chinese Academy of Sciences, Wuhan 430071, China}
\affiliation{Hefei National Laboratory, Hefei 230088, China}

\author{Jin Wang}
\affiliation{Innovation Academy for Precision Measurement Science and Technology, Chinese Academy of Sciences, Wuhan 430071, China}
\affiliation{Hefei National Laboratory, Hefei 230088, China}
\affiliation{Wuhan Institute of Quantum Technology, Wuhan 430206, China}

\author{Ming-Sheng Zhan}
\email{mszhan@wipm.ac.cn}
\affiliation{Innovation Academy for Precision Measurement Science and Technology, Chinese Academy of Sciences, Wuhan 430071, China}
\affiliation{Hefei National Laboratory, Hefei 230088, China}
\affiliation{Wuhan Institute of Quantum Technology, Wuhan 430206, China}

\begin{abstract}
Weak equivalence principle tests comparing different elements may be more sensitive to violations than those comparing isotopes. However, differences in effective wave vector complicate common-mode rejection of vibration noise. We propose a Moir\'e-enabled recursive phase unwrapping (MRPU) method for differential measurements with dual-species atom interferometers. The method exploits the moir\'e structure of the joint phase likelihood and temporal acceleration correlations to unwrap both phases and determine the relative acceleration. We implemented an $^{85}$Rb $\mathrm{D_1/D_2}$ line dual-scale-factor interferometer with an identical normalized transfer function and a large effective wave-vector difference to emulate different atomic species. Under $40\,\mathrm{rad}$ equivalent vibrational noise, this method achieved phase unwrapping over 5000 consecutive measurements, with a relative-acceleration Allan deviation of $1.47\times10^{-8}g$ at 997 shots. Numerical simulations under random vibrations and slow drifts show a $9.76$--$26.03\,\mathrm{dB}$ increase in the tolerable narrowband vibration amplitude. This method is promising for weak equivalence principle tests based on different atomic species, dark matter and dark energy detection, and high-dynamic-range atomic gravimetry.
\end{abstract}

\maketitle

\textit{Introduction}---The weak equivalence principle (WEP) states that,  in the same gravitational field, the acceleration of a freely falling test body is independent of its composition or internal structure. Torsion balances, lunar laser ranging, and the MICROSCOPE satellite have tested this principle with high precision \cite{Schlamminger_2008,Williams_2004,Touboul_2022}. Light-pulse atom interferometers provide a microscopic test of WEP by comparing the free-fall accelerations of different elements, isotopes, or internal states \cite{Fray_2004,Schlippert_2014}.

Following the first $^{85}\mathrm{Rb}$ and $^{87}\mathrm{Rb}$
atom-interferometric WEP test at the $10^{-7}$ level \cite{Fray_2004} in 2004, the precision of Rb-isotope WEP tests improved only slowly over more than a decade. As long-baseline atom interferometers \cite{Dickerson_2013} have gradually been combined with ultracold atomic sources \cite{Kovachy_2015}, high common-mode rejection differential measurements \cite{Bonnin_2013,Zhou_2015}, large momentum transfer \cite{Chiow_2011}, rotation compensation \cite{Lan_2012}, phase-shear readout \cite{Sugarbaker_2013}, and gravity-gradient compensation \cite{Roura_2017,Overstreet_2018}, the uncertainty of Rb-isotope WEP tests has been reduced to the $10^{-12}$ level \cite{Asenbaum_2020}.

Differences in mass and internal composition among atoms of different elements are generally much greater than those among isotopes, and many new theories predict that such combinations may be more sensitive to effects arising from a violation of WEP \cite{Hohensee_2013,Schlippert_2014}. However, the effective wave vectors of different atomic species vary significantly, and the phase shift caused by vibration noise can easily exceed \(2\pi\), making it difficult to extract the differential signal. Traditional fringe fitting typically requires efficient suppression of vibrational noise\cite{Schlippert_2014, Albers_2020}. To extract the phase in the presence of strong vibrational noise, several approaches have been proposed: (i) combining different free-evolution times \(T\) to achieve common-mode noise suppression under similar scale factors and extracting the phase shift with ellipse fitting\cite{Barrett_2016}; or (ii) extending the measurement range of atom interferometers and using classical accelerometers for phase unwrapping\cite{Barrett_2015, Barrett_2022}. These approaches have demonstrated effectiveness under the relevant experimental conditions; however, for higher precision, the consistency of the normalized transfer function and nonlinear effects in classical accelerometers still require further attention. Previous studies have used moir\'e methods to extend the measurement range without auxiliary accelerometers, primarily in single-species atom interferometers \cite{Avinadav_2020,Yankelev_2020}.

In this Letter, we propose the MRPU method for differential measurements using dual-species atom interferometers with different elements and demonstrate it experimentally. The method exploits the moir\'e structure of the joint phase likelihood and temporal acceleration correlations to unwrap both phases and determine the relative acceleration. To validate this method, we designed and implemented a $^{85}\mathrm{Rb}$ $\mathrm{D_1/D_2}$ line dual-scale-factor atom interferometer. The two interferometers have matched normalized vibration responses. Their effective-wave-vector difference emulates the scale-factor difference between interferometers using different elements, providing an experimental test of MRPU. Numerical simulations further evaluate the method under random vibration and slow drift.

\textit{Methods}---Consider two interferometers with scale factors $S_i=k_iT^2$ and phases $\Phi_{i,n}=S_i g_{i,n}$ at measurement $n$. Here, $k_i$ is the effective wave vector, and known laser-phase offsets are omitted. The acceleration $g_{i,n}$ includes the common vibration contribution. Fringe fitting gives the measured phase $\phi_{i,n}\in[-\pi,\pi)$. Define $\bar g=(g_1+g_2)/2$, $\delta g=(g_1-g_2)/2$, $\bar S=(S_1+S_2)/2$, and $\delta S=(S_1-S_2)/2$. Thus, the relative acceleration is $g_1-g_2=2\delta g$. For identical pulse timing,
\begin{equation}
\Phi_d\equiv\Phi_1-\frac{S_1}{S_2}\Phi_2=S_1(g_1-g_2)=2S_1\delta g.
\label{eq:differential}
\end{equation}
The common vibration contribution cancels in $\Phi_d$ \cite{Chen_2014,Barrett_2022}. Directly replacing $\Phi_i$ by $\phi_i$ generally fails once either phase leaves its $2\pi$ measurement range. The normalized vibration responses are matched by using identical pulse timing and Rabi frequencies \cite{Barrett_2015}.

MRPU uses the two measured phases to estimate the average acceleration and select the appropriate $2\pi$ intervals. Assume independent phase-readout noise with the same von Mises distribution. Define $\bar\phi_n=(\phi_{1,n}+\phi_{2,n})/2$ and $\delta\phi_n=(\phi_{1,n}-\phi_{2,n})/2$. Up to an additive constant, the joint log likelihood is
\begin{equation}
\ln\mathcal L_n\propto\cos(\bar\phi_n-\bar S\bar g-\delta S\delta g)\cos(\delta\phi_n-\delta S\bar g-\bar S\delta g).
\label{eq:likelihood}
\end{equation}
In our experiment, variations in $\delta g$ are much smaller than those in $\bar g$. As $\bar g$ varies, the two cosine factors have acceleration periods $2\pi/\bar S$ and $2\pi/|\delta S|$. The slow factor produces a Moir\'e envelope that extends the measurement range and gives a coarse estimate of $\bar g$, while the fast factor provides closely spaced likelihood peaks for determining it with higher resolution.

Differential phase noise can shift the coarse estimate toward a neighboring likelihood peak and lead to errors of approximately $2\pi$ in the unwrapped phases. MRPU reduces these errors through Bayesian recursion. Let $P_n(\mathbf g_n)$ be the posterior after $n$ measurements, with $\mathbf g_n=(\bar g_n,\delta g_n)$. Then
\begin{equation}
P_n\propto\mathcal L_n\int K_n(\mathbf g_n\mid\mathbf g_{n-1})P_{n-1}\,d\mathbf g_{n-1}.
\label{eq:bayes}
\end{equation}
The transition kernel $K_n$ describes the expected acceleration changes between measurements. It propagates the previous posterior into a predictive prior, which is then multiplied by the current likelihood. Combining this temporal prediction with the measured phases suppresses erroneous selection of neighboring likelihood peaks. MRPU uses the estimated average acceleration to calculate the unwrapped phases $\widehat\Phi_{1,n}$ and $\widehat\Phi_{2,n}$. It then uses Eq.~\eqref{eq:differential} to calculate the differential phase $\phi_{d,n}^{\mathrm{MRPU}}$ and obtain the relative acceleration.

\textit{Experimental setup}---We operate synchronous $^{85}\mathrm{Rb}$ interferometers on the $D_1$ ($794.98\,\mathrm{nm}$) and $D_2$ ($780.24\,\mathrm{nm}$) lines, with spatial phase-shear readout. Their effective wave vectors satisfy $|k_1-k_2|\approx3.0\times10^5\,\mathrm{m}^{-1}$ and $\bar k=(k_1+k_2)/2\approx1.60\times10^7\,\mathrm{m}^{-1}$. The relative difference is $|k_1-k_2|/\bar k\approx1.9\%$, which is comparable to that of Rb ($780\,\mathrm{nm}$) -- K ($767\,\mathrm{nm}$) atom interferometers. Using the same isotope allows the two interferometers to share atomic preparation, pulse control, and imaging, reducing preparation and detection differences in this method test.

\begin{figure}[tbp]
\centering
\includegraphics[width=\linewidth]{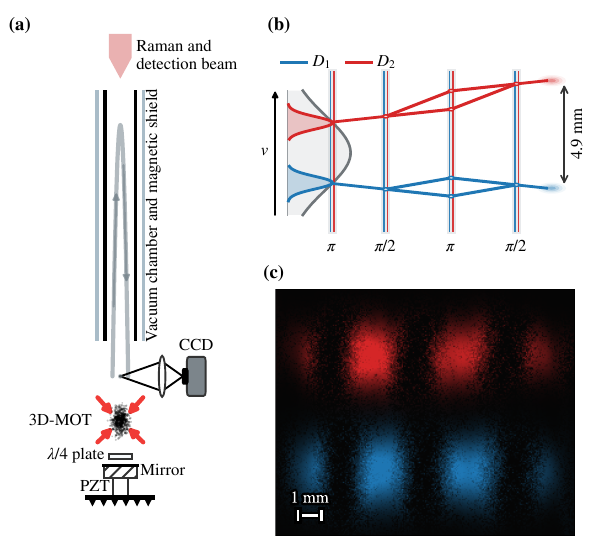}
\caption{Synchronous $D_1/D_2$ line atom interferometers. (a) Experimental apparatus. (b) $D_1$ and $D_2$ Raman pulses select atoms near velocities $-v$ and $+v$, respectively. The two ensembles then undergo synchronous $\pi/2$--$\pi$--$\pi/2$ sequences. (c) Spatial phase-shear fringes recorded in one CCD image. The two clouds are separated by $4.9\,\mathrm{mm}$.}
\label{fig:setup}
\end{figure}

The apparatus is shown in Fig.~\ref{fig:setup}(a) \cite{Yan_2023}. A two-dimensional magneto-optical trap loads approximately $10^9$ atoms into a three-dimensional trap in $1.2\,\mathrm{s}$. Moving optical molasses and polarization-gradient cooling prepare an upward-launched cloud at a few microkelvin, with a launch velocity of approximately $3\,\mathrm{m/s}$. In a magnetically shielded region, the two Raman pairs drive synchronous vertical $\pi/2$--$\pi$--$\pi/2$ Mach--Zehnder sequences with $T=50\,\mathrm{ms}$. The atoms are detected approximately $17\,\mathrm{cm}$ above the three-dimensional trap using a vertical probe on the $|F=3\rangle\rightarrow|F'=4\rangle$ transition and CCD imaging.

The Raman pairs originate from independent $795$- and $780$-nm seed lasers. After frequency shifting by approximately $\pm1.5\,\mathrm{GHz}$, they are combined by a dichroic mirror and pulsed by a common $110\,\mathrm{MHz}$ acousto-optic modulator. The combined beams enter the apparatus through a polarization-maintaining fiber and are reflected by a mirror at the bottom. Double passage through a quarter-wave plate produces Doppler-sensitive counterpropagating Raman beams. Both Raman pairs are red-detuned by approximately $1.5\,\mathrm{GHz}$ from their respective optical transitions. Their intensity ratios are $3.6{:}1.0$ and $3.7{:}1.0$ for $D_1$ and $D_2$, respectively, to reduce differential ac Stark shifts.

Velocity-selection pulses with two-photon detunings of approximately $-30\,\mathrm{kHz}$ and $+30\,\mathrm{kHz}$ prepare two ensembles in $|F=2,m_F=0\rangle$ (Fig.~\ref{fig:setup}(b)). The momentum-transfer directions are opposite during selection but the same during interferometry. The selected ensembles differ in Raman resonance frequency by approximately $60\,\mathrm{kHz}$. The two Raman pairs have matched Rabi frequencies. Their $100\,\mu\mathrm{s}$ $\pi$ pulses have a transition linewidth of approximately $5\,\mathrm{kHz}$, which suppresses excitation by the other Raman pair. The clouds are separated by approximately $4.9\,\mathrm{mm}$ at detection.

Before the final Raman pulse, a piezoelectric actuator tilts the mirror by approximately $120\,\mu\mathrm{rad}$. This creates a spatial phase gradient across each ensemble \cite{Sugarbaker_2013,Yan_2023}. One CCD image records both fringe patterns (Fig.~\ref{fig:setup}(c)). Separate sinusoidal fits yield $\phi_{1,n}$ and $\phi_{2,n}$ without scanning the interferometer phase for readout.

To test phase unwrapping beyond the $2\pi$ measurement range, we scan the two Raman chirp rates in proportion to their wave vectors \cite{Chen_2014}. The equivalent acceleration changes, $\Delta a_i=2\pi\Delta\alpha_i/k_i$, are therefore equal, where $\alpha_i$ is the Raman frequency chirp rate. Phase unwrapping can accommodate larger phase excursions, as shown by the simulations below. To avoid excessive Raman detuning and loss of signal in this test, we reverse the scan at a phase excursion of approximately $40\,\mathrm{rad}$, equivalent to approximately $10^{-3}\,\mathrm{m/s^2}$. Each scan period contains 200 measurements, and 5000 measurements are acquired. This controlled scan provides a vibration-equivalent test of phase unwrapping.

\begin{figure}[tbp]
\centering
\includegraphics[width=\linewidth]{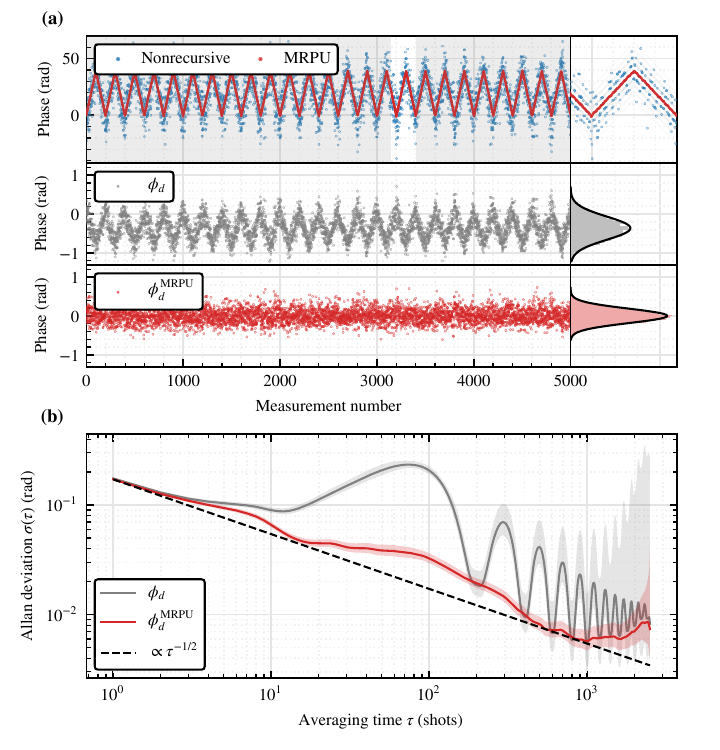}
\caption{Phase unwrapping and differential measurement. (a) From top to bottom: $\bar S\bar g$ for 5000 measurements and an enlarged view, the differential phase $\phi_d$ calculated from the measured phases and its distribution, and the differential phase calculated from the unwrapped phases, $\phi_d^{\mathrm{MRPU}}$, and its distribution. Blue and red points show single-measurement likelihood estimates and recursive estimates, respectively. (b) Allan deviations of $\phi_d$ (gray) and $\phi_d^{\mathrm{MRPU}}$ (red), with confidence bands. The averaging interval is given in numbers of measurements.}
\label{fig:phase_results}
\end{figure}

\textit{Results and simulations}---Figure~\ref{fig:phase_results}(a) shows the average-acceleration phase $\bar S\bar g$ over the measurements. Blue points are likelihood estimates from individual measurements, and red points include Bayesian recursion. The single-measurement estimates select different phase intervals from the recursive result in $74.98\%$ of measurements. With temporal correlations included, the unwrapped phase follows the applied chirp scan throughout the sequence.

The differential phase calculated from the measured phases, $\phi_{d,n}=\phi_{1,n}-(S_1/S_2)\phi_{2,n}$, varies periodically with the scan. MRPU unwraps the measured phases to obtain $\widehat\Phi_{1,n}$ and $\widehat\Phi_{2,n}$. The differential phase calculated from these unwrapped phases, $\phi_{d,n}^{\mathrm{MRPU}}$, removes this periodic variation. Its standard deviation is $0.184\,\mathrm{rad}$, corresponding to $4.67\times10^{-6}\,\mathrm{m/s^2}$ for the relative acceleration. The large chirp scan can increase phase-readout noise by moving the Raman pulses away from resonance.

The Allan deviation of $\phi_d$ has minima at averaging intervals equal to integer multiples of the 200-measurement scan period (Fig.~\ref{fig:phase_results}(b)). At these intervals, averaging suppresses the periodic scan contribution, revealing the apparatus's differential noise floor. Random vibration does not undergo this periodic cancellation, leaving a residual vibration contribution that can increase the Allan deviation. In contrast, the Allan deviation of $\phi_d^{\mathrm{MRPU}}$ decreases approximately as $\tau^{-1/2}$ and follows the lower envelope of the $\phi_{d,n}$ curve. Agreement with this envelope supports recovery of the apparatus's differential sensitivity without artificially suppressing the measured fluctuations. The Allan deviation of $\phi_{d,n}^{\mathrm{MRPU}}$ reaches $5.70\times10^{-3}\,\mathrm{rad}$ at an averaging interval of 997 measurements, equivalent to $1.47\times10^{-8}g$ in relative acceleration.

We next simulate random vibration and slow drift using the same pulse separation and relative scale-factor difference as in the experiment. The input is narrowband Gaussian random vibration with quality factor $Q=10$. It is filtered by the three-pulse interferometer's acceleration transfer function and sampled every $2\,\mathrm{s}$. We add independent phase noise of $0.02\,\mathrm{rad}$ standard deviation to each interferometer and include a small sinusoidal differential phase.

\begin{figure}[tbp]
\centering
\includegraphics[width=\linewidth]{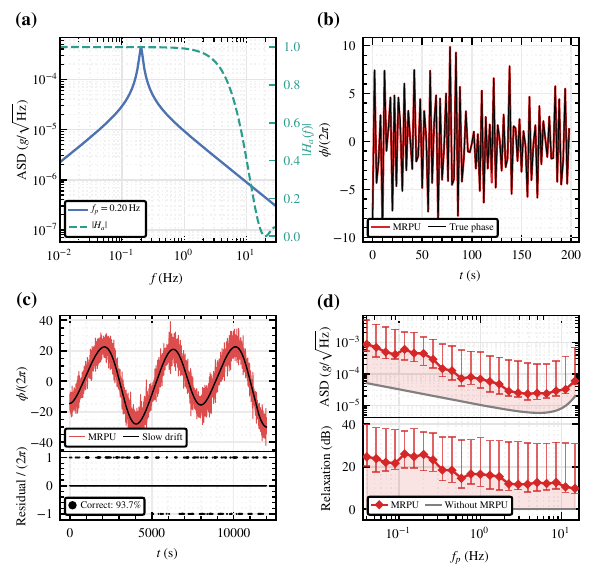}
\caption{Simulated MRPU performance under vibration and slow drift. (a) Input vibration ASD and the magnitude of the acceleration transfer function. (b) Phase unwrapping at $f_p=0.20\,\mathrm{Hz}$ and peak ASD $4.5\times10^{-4}g/\sqrt{\mathrm{Hz}}$. The success rate is $93.8\%$. (c) Unwrapped phases and residuals after adding slow drift (black curve). The success rate is $93.7\%$. (d) Peak-ASD thresholds for $95\%$ phase unwrapping success (red) and the same probability of remaining within $[-\pi,\pi)$ without phase unwrapping (gray). Below: $20\log_{10}$ of the threshold ratio, with $95\%$ confidence intervals.}
\label{fig:simulation}
\end{figure}

For a vibration spectrum centered at $f_p=0.20\,\mathrm{Hz}$ with peak amplitude spectral density (ASD) $4.5\times10^{-4}g/\sqrt{\mathrm{Hz}}$, the average-acceleration phase spans approximately $24\times2\pi$ peak to peak. The phase unwrapping success rate is $93.8\%$ (Fig.~\ref{fig:simulation}(b)). Adding a slow drift extends the total phase range to approximately $81\times2\pi$, with a success rate of $93.7\%$ (Fig.~\ref{fig:simulation}(c)).

We vary $f_p$ from $0.04$ to $14.62\,\mathrm{Hz}$ and the input vibration amplitude. Figure~\ref{fig:simulation}(d) compares two thresholds: the peak ASD at which MRPU achieves a mean unwrapping success rate of $95\%$, and that at which the vibration-induced phase remains within its $2\pi$ measurement range with the same probability. The MRPU threshold is higher by $9.76$--$26.03\,\mathrm{dB}$. These simulations show an increased tolerance to vibration.

\textit{Conclusion}---In summary, we proposed a Moir\'e-enabled recursive phase unwrapping method for differential measurements with dual-species atom interferometers based on different elements, and developed a synchronized $^{85}\mathrm{Rb}$ $\mathrm{D_1/D_2}$ line dual atom interferometer to validate it. Experimental results show that, under an equivalent vibrational phase shift of $40\,\mathrm{rad}$, the MRPU method achieves phase unwrapping over 5000 consecutive measurements, and the Allan deviation of the differential gravity measurement is $1.47\times10^{-8}g$ at 997 shots. Further numerical simulations show that, under random vibrations and slow drifts, the method increases the tolerable narrowband vibration amplitude by $9.76$--$26.03\,\mathrm{dB}$.

Unlike conventional classical-accelerometer-assisted correlation measurement methods \cite{Barrett_2022,Barrett_2015}, the MRPU method requires no auxiliary classical accelerometer and thus avoids limitations such as sensor nonlinearity. Moreover, because its Moir\'e effect arises from the wave vector difference, the two interferometers share the same normalized transfer function, offering higher ultimate measurement precision than modifying \(T\) to equalize the scale factors. Combined with vibration isolation and long-baseline atom interferometry, MRPU should enable differential measurements with dual-species interferometers of different elements to approach or match isotope dual-species performance, thereby improving the precision of WEP tests \cite{Schlippert_2014,Barrett_2022}. It is also promising for dark-matter and dark-energy detection with dual-species interferometers using different element combinations and for dynamic gravity measurements.

\textit{Acknowledgments}---We thank Yu-Hang Ji and Yan Wang for their contributions to the apparatus and related techniques. This work was supported by the National Natural Science Foundation of China (Grant Nos. U25D9001, W2412045, 12174403, and U25D8014),the National Innovation Program for Quantum Science and Technology of China (Grant Nos. 2021ZD0300603,2021ZD0300604), the Chinese Academy of Sciences Project for Young Scientists in Basic Research (Grant No. YSBR-055), and the Natural Science Foundation of Hubei Province (Grant No. 2022CFA096).

\bibliography{Ref}
\end{document}